# Activation energies of water molecular dynamics and Stokes-Einstein equation

Alexander Kholmanskiy

Science Center «Bemcom», Moscow, Russia

allexhol@ya.ru, http://orcid.org/0000-0001-8738-0189

**Abstract**

An analysis of the values and signs of the activation energies of temperature dependences (TDs) of the self-diffusion coefficient (D) and dynamic viscosity (η) in the range from 0 °C to 100 °C confirmed that the molecular dynamics of water is based on the synergism of endo and exo reactions of rearrangement of hydrogen bonds in the supramolecular structure water. The ratio of linear approximation coefficients TD of the complex characteristic Dη before and after 25 °C correlates with the ratio of cluster sizes prevailing in the water structure in these temperature ranges. This result is consistent with the Stokes-Einstein equation and the hypothesis that the decomposition of hexagonal ice-like clusters is completed in the vicinity of 25 °C. A reliable approximation of the TD complex Dη by a bimodal function of type *Texp (E/RT)* was used to refine the Stokes-Einstein equation.

## 1. Introduction

In [1], it was established that the molecular physics of the anomalous temperature dependences (TDs) of water characteristics is based on the synergy of endo and exo reactions of rearrangement of the network of dynamic hydrogen bonds (HBs) in the supramolecular structure (SMS) of water. Moreover, TDs can be reliably approximated by the bimodal temperature function (FA) of the following form:

$$TD = F_A = T^{\pm\beta} F_R = \exp(\pm E_A/RT), \qquad (1)$$

$\beta = 0, 1, \frac{1}{2}, 2$, and $F_R$ approximates the modified TD*:

$$TD^* = F_R = TD/T^{\pm\beta} = \exp(\pm E_R/RT),$$

$+E_R$ corresponds to the activation energy of the exo, and $-E_R$ endo reaction of the restructuring of the water structure. Approximating $T^{\pm\beta}$ by the function $exp\,(\pm E_T/RT)$, we obtained:

$$F_A = F_T F_R = \exp(\pm E_T/RT) \cdot \exp(\pm E_R/RT) = \exp(\pm E_A/RT),$$

$$\pm E_A = \pm E_R \pm E_T.$$

At the extreme points TDs of density and specific volume (4 °C), isobaric heat capacity (35 °C), compressibility (46 °C), and sound velocity (75 °C), the values of $E_R$ and $E_T$ are equal in absolute value and opposite in sign, therefore $E_A$ values at the extreme points are equal to zero, and to the right and left of them have different signs [1].

The trends of $F_A$ approximations TDs of the self-diffusion coefficient (D), spin-lattice relaxation time ($T_1$), and dynamic viscosity (η) have a kink in the vicinity of 25 °C [1, 2]. Taking into account the Stokes-Einstein equation:

$$D = \frac{k_B T}{6\pi \eta r}, \qquad (2)$$

($k_B$ – the Boltzmann constant, **r** is the particle radius) for D and $T_1$ in [1, 2] took β = 1, and for η β = 0. $E_A$ values for these characteristics were comparable with the hydrogen bond energy and varied greatly in the intervals 0 -25 °C and 26-100 °C. Moreover, in the case of D and $T_1$, the $E_A$ values were negative (endo reaction), and for η, positive (exo reaction). In [1, 2, 3] it was suggested that, in the vicinity of 25 °C, the transformation of the ice-like supramolecular structure is completed by the decomposition of tetrahedral hexagonal clusters ($W_6$) into dimers and smaller clusters.

As a rule, the value D is calculated from $T_1$ values, which are determined by the spin-lattice relaxation of protons, measured by the NMR. A confirmation of the adequacy of the D values is usually the relation (2) [4, 5, 6]. By definition, D should characterize the displacement, that is, the translational mobility of the center of gravity of the water molecules located next to the oxygen atom. The dynamics of proton spins in the general case includes rotational motions of molecules, tunneling jumps of a proton along HBs chains and differs for spin ortho and para isomers of water [7]. The ambiguity of the correspondence between the dynamics of proton spins and the mobility of oxygen atoms is expressed in the qualitative nature of the correlations between the values of D and $T_1$. Equation (2) is also evaluative, since it was derived using a statistical model of thermodynamics that does not take into account the TDs features of the dynamic characteristics of water.

The bimodal function (1) was used for the analysis of TDs D, $T_1$ and $\eta$ in order to check the adequacy of the Stokes-Einstein equation and determine the activation energies of the reactions responsible for the rearrangement of the water structure in the vicinity of 25 °C.

## 2. Results and discussion

$F_A$, $F_R$ and $F_T$ approximations of TDs of characteristics D, $T_1$ and $\eta$, as well as their compositions $D\eta$ and $D/T_1$, were performed similarly [1]. For D, $T_1$ and $D\eta$, the $F_A$ functions were of the form $T\exp(-E_A/RT)$, and for $\eta$ and $D/T_1$, $\exp(E_A/RT)$ and $\exp(-E_A/RT)$, respectively. $R$ is the gas constant (8.3, J · mol$^{-1}$ · K$^{-1}$). To obtain the activation energy values in J mol$^{-1}$, the numerical coefficients of the exponential trend indicators were multiplied by $R$. The variation of the values of D and $T_1$ obtained in different works could reach ±20-30% [5, 8], so the values of D and $T_1$ were averaged and indicated deviation range. The $\eta$ values in [9, 10] coincided for almost all T. Figure 1 shows the dependences of $D\eta$ on T and 1/T, as well as their approximations. The activation energies $E_A$, $E_R$, and $E_T$ are shown in the Table.

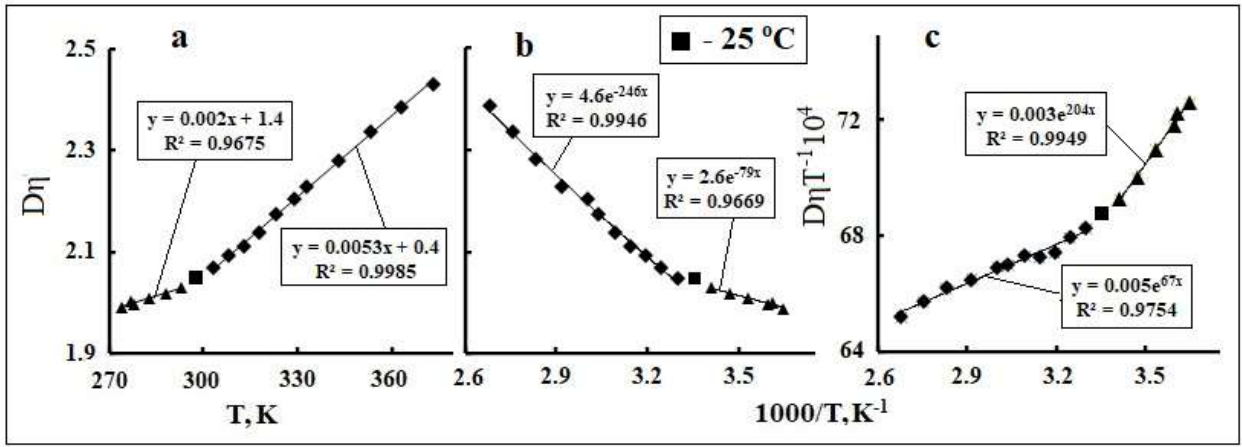

**Figure 1**. (**a**) Dependence of $D\eta$ on T and its linear approximation; (**b**) the dependence of $D\eta$ on 1/T and its $F_A$ approximation; (**c**) dependence of $D\eta T^{-1}$ on 1/T and its $F_R$ approximation. The black square marks the extreme point of 25 °C. Initial data from [9, 10, 11]

From Figure 1 it follows that TDs $D\eta$ and their approximations have a kink in the vicinity of 25 °C, which is typical for all dynamic characteristics of water (see Table and [1, 2, 3]). The ratio of linear trend coefficients TD $D\eta$ in the intervals of 0-25 °C and 26-100 °C is equal to ~2.7 (see Figure 1a). Since in accordance with (2) these coefficients are $k_B/6\pi r$, the parameter $r$ at T < 25 °C should be 2.7 times larger than at T > 25 °C. This result is consistent with the hypothesis of a sharp transformation of SMS in the vicinity of 25 °C and the decay of $W_6$ into dimers and clusters that are 2-3 times smaller than the size of $W_6$ [1, 4]. Transformation also corresponds to the transition between two metastable phases of SMS water – low density liquid (LDL) and high density liquid (HDL) [1]. The prevailing loose tetrahedral configuration of HBs is characteristic of LDL, and the denser configuration of disturbed HBs is predominant for HDL [1].

**Table**

Temperature interval, parameter β and activation energy of Arrhenius approximations of the temperature dependences of the water characteristics.

| Water characteristics | β | ΔT (°C) | $E_T$ | $E_R$ | $E_A$ ($E_R + E_T$) | [Ref] |
|---|---|---|---|---|---|---|
| | | | | kJ/mol | | |
| D (cm² s⁻¹) | 1 | 0 – 25 | -2.5 | -18.4±0.4 | -20.9±0.3 | [2, 6, 11] |
| | | 26-100 | -2.8 | -14.1±0.1 | -16.9±0.1 | |
| $T_1$(s) | | 0 – 25 | -2.5 | -18.5±0.7 | -21±0.5 | [2, 11, 12] |
| | | 30-100 | -2.8 | -12.5±0.6 | -15.3±0.6 | |
| η (cP) | 0 | 1 – 25 | 0 | | 18.6 | [9, 10] |
| | | 26 – 100 | | | 14.0 | |
| D/$T_1$ | 0 | 0 – 25 | | | -0.8 | [2, 6, 11] |
| | | 30-100 | | | -2.8 | |
| Dη | 1 | 1 – 20 | -2.7 | 1.7 | -0.7 | [9, 11] |
| | | 25 – 100 | -2.6 | 0.6 | -2.0 | |

The values of $E_A$ and $E_R$ for D and $T_1$ practically coincide in the range of 0-25 °C, but in the range of 26-100 °C their values for D are 10-12% higher than the values for $T_1$. Moreover, the difference in the values obtained by different authors does not exceed ~5% (see Table). This result indicates an increase in the mismatch in the physics of molecular displacement and reorientation of proton spins with an increase in T. The facilitation of the latter process in the range of 26-100 °C is apparently due to an increase in the rotational degrees of freedom of the molecules due to a decrease in the density of HBs in SMS.

Taking into account the coincidence of the absolute values of $E_R^D$ and $-E_A^\eta$, for D and η, in the framework of the theory of the activated complex, we constructed a diagram of the energy levels of the water molecule involved in the self-diffusion process (Figure 2). According to the scheme, the instantaneous self-diffusion act is the conjugation of the endothermic stage of rupture of existing HBs with the exothermic stage of the formation of new HBs. The energy $E_A^\eta$ released in this case is used to initiate the process of displacement of other molecules. The rearrangement of HBs is accompanied by a change in the environment of the molecule and its displacement in space. The first stage is responsible for the physics of viscosity, and the physics of the second stage is related to the physics of the process of water crystallization [1]. The kinetics of both stages depends on the instantaneous structure of the hydrogen bond network in SMS, which in turn depends on T.

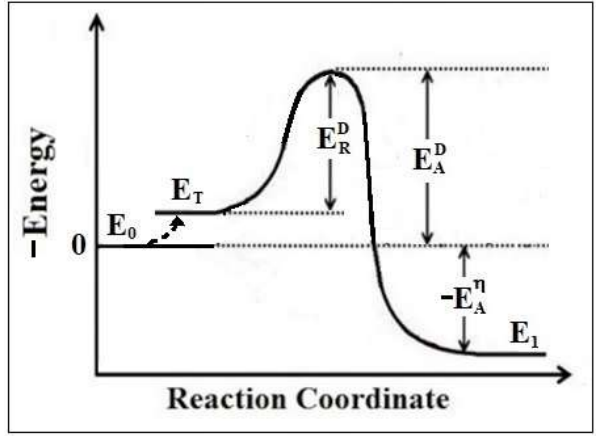

Figure 2. Energy profiles of endo- ($E_R^D$) and exothermic ($-E_A^\eta$) stages of the process of water self-diffusion. $E_0$ and $E_1$ energy of the initial and final state of the molecule; $E_T$ is the activation energy of the thermal component of the process.

The proximity of $E_A$ values for compositions Dη and D/T₁ indicates the limitation of their TDs by the activation mechanism of the endothermic stage of the self-diffusion process. The kinetics of this mechanism is determined by the activation energy $E_R^{D\eta}$, which is 2.8 times greater in the range of 0-25 °C than in the range of 25-100 °C (see Table). This result can be associated with a corresponding increase in the average number of HBs per water molecule in the range of 0–25 °C due to the predominance of $W_6$ clusters with tetrahedral HBs in SMS [1].

Within the bounds of reliability of the approximation of TDs of the complex Dη by the function $exp(\frac{E_R^{D\eta}}{RT})$ (Figure 1c), the Stokes-Einstein equation can be represented as:

$$D = \frac{k_B T}{6\pi\eta r} \exp\left(\frac{E_R^{D\eta}}{RT}\right),$$

where $E_R^{D\eta}$ in the intervals 0-25 °C and 26-100 °C is 1.7 kJ/mol and 0.6 kJ/mol, respectively.

3. Conclusion

An analysis of the temperature dependence of the composition of the self-diffusion coefficient and the dynamic viscosity of water (Dη) confirmed the hypothesis that the decomposition of ice-like hexagonal clusters into smaller clusters in the vicinity of 25 °C is completed. The ratio of their sizes qualitatively correlates with the ratio of the radii appearing in the Stokes-Einstein equation, as well as the coefficients of a linear approximation of the dependence of Dη on temperature. The proximity of the values and the difference in the signs of the activation energies obtained from the Arrhenius approximations TDs D and η were used to construct the energy profile of the self-diffusion process. The bimodal approximation parameters TD of the composition Dη made it possible to refine the form of the Stokes-Einstein equation.